\documentclass{ws}

\begin{document}

\title{{\noindent\small \rm UNITU-THEP-11/01
          \hfill hep-ph/0105320} \\ 
Nucleon Form Factors in the Covariant Diquark--Quark Model\thanks{
\small \rm Talk given by
R.A.\ at the  Workshop on {\em Lepton scattering, Hadrons and QCD}, Adelaide, 
March 26 to April 6, 2001.}}

\author{R.\ Alkofer}

\address{Institute for Theoretical Physics, T\"ubingen University,\\
Auf der Morgenstelle 14, D-72076 T\"ubingen, Germany,\\ 
E-mail: Reinhard.Alkofer@uni-tuebingen.de}

\author{M.\ Oettel}

\address{CSSM, University of Adelaide, 10 Pulteney St, Adelaide, SA, 5005\\
E-mail: moettel@physics.adelaide.edu.au}

\maketitle

\abstracts{
The nucleon is described as a bound state of a quark and an extended diquark.
Hereby the notion ``diquark'' refers to the modelling of separable correlations
in the  two--quark Green's functions. Binding of quarks and diquarks takes
place via an exchange interaction and is therefore related to the Pauli
principle for three--quark states. Fully Poincar{\'e} covariant nucleon
amplitudes are calculated for free constituent propagators as well as for 
dressed propagators which parameterise confinement.  The corresponding results
for space--like form factors differ quantitatively but not qualitatively for
various {\it ans\"atze} for the propagators.  These results do not allow to
draw definite conclusions on the  permissibility  of different dressing
functions. Results for kaon  photoproduction, on the  other hand, exclude a
whole class of constituent  propagators.}

\section{Motivation}

The central aim of the studies reported here is to develop a phenomenological 
QCD--based model for baryon structure. Recent experimental results emphasise
the complicated nature of baryons. An example is provided by the ratio
of the electric to the magnetic form factor of the proton measured  
at Jefferson Lab\cite{Jones:2000rz}: This ratio surprisingly
decreases with increasing photon virtuality. We will see later on how this can
be understood in our model.

Theoretical issues such as confinement, dynamical breaking of chiral symmetry
and the formation of relativistic bound states can be understood and related to
the properties of the non-perturbative propagators of QCD. In the
Dyson--Schwinger approach we have obtained remarkable progress on this
interrelation and other kindred questions during the last years, see the recent
reviews\cite{Alk00,Rob00}. Amongst many important results we want to highlight 
the following: Using the general structure of the ghost Dyson--Schwinger
equation in Landau gauge, Slavnov--Taylor identities and multiplicative
renormalisibility one only needs the assumption that QCD Green's functions can
be expanded in asymptotic series in the infrared to demonstrate confinement
of transverse gluons\cite{Wat01,Alk01}. Especially, the infrared behaviour of
the gluon and ghost propagators are uniquely related. This prediction can be
tested in future lattice calculations as e.g.\ the ones reported in 
ref.\cite{Bon00}.

Corresponding studies of quark confinement are under way, and lattice results
for the quark propagator\cite{Skullerud:2000un} will serve to guide them. Since
results are not available yet, and because studies of baryon properties require to
avoid unphysical thresholds a pragmatic way to proceed is via the
parameterisation of quark propagators, i.e., by multiplying dressing functions
to the quark propagators: 
\begin{eqnarray} 
S^{(k)}(p) &=&  \frac{i{p  \!\!\! /}-m_{q}}{p^2+m_{q}^2} 
{f_k\left( p^2/{m_{q}^2} 
\right)} . 
\end{eqnarray} 
The trivial dressing function
$f_0\equiv 1$ corresponds to the bare propagator.  ``Confining'' ans\"atze 
include\cite{Ahlig:2000qu}
\begin{eqnarray} 
f_1 (x) &=& \frac{1}{2}\left\{ \frac{x+1}{x+1-i/d} +
\frac{x+1}{x+1+i/d} \right\} \label{f1} \; ,\\ f_2 (x) &=& 1- \exp\left[
-d\left(1+x \right) \right] \label{f2}\; , \\ f_3 (x,x^\ast) &=& \tanh\left[
d\left(1+x\right) \left(1+x^\ast\right) \right] \; . \label{f3} 
\end{eqnarray}
The dressing functions $f_k$ behave qualitatively different for time-like 
momenta:
${f}_1(x)$ and ${f}_3(x)$ change sign (as in the case of a
tree--level propagator) and the function ${f}_2(x)$, that models the
exponential type propagator, increases drastically. Away from the real axis,
the behaviour of the above model propagators can be read off easily from the
definitions (\ref{f1})--(\ref{f3}). By construction the propagators $S^{(1)}$
have complex conjugated poles at $p^2 = m^2(-1\pm i /d)$ where
$m$ represents a parameter that would be interpreted as a mass if and only if
the poles were on the real axis. The entire propagator $S^{(2)}$
oscillates for a nonzero imaginary part of $p^2$. The propagator $S^{(3)}$ is
built such that its dressing function $f_3$ asymptotically approaches unity in
all directions of the complex $p^2$ plane. For asymptotically large space-like
momenta the three model propagators $S^{(k)},\, (k=1,2,3)$ match up with the
bare propagator $S^{(0)}$.

\begin{figure}[t]
\epsfxsize=0.8\linewidth
\centerline{\epsfbox{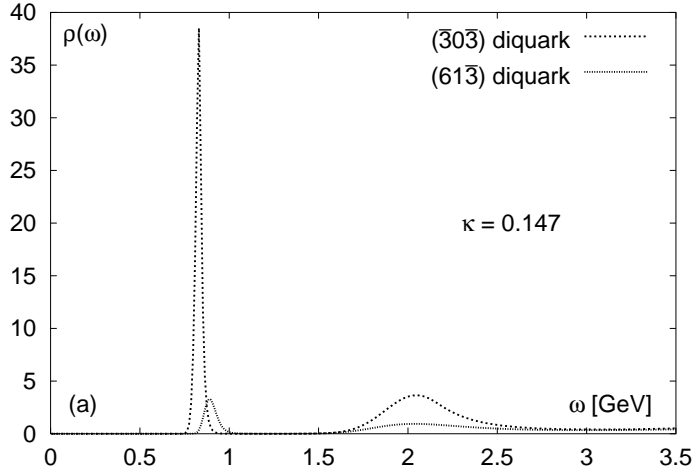} }
\caption{Spectral functions for diquarks from lattice data.
$\bar 30 \bar 3$ denotes scalar diquarks, $61\bar 3$ axialvector ones. 
(Adopted from ref.\protect\cite{Wetzorke:2000ez}.) 
\label{diq}}
\end{figure}

Diquarks appear in many phenomenological models. The quantum numbers of
one--gluon--exchange suggest that not only colour singlet $\bar qq$ pairs,
i.e., mesons, are bound but also colour triplet $qq$ pairs. Closer analysis
shows though that in model Bethe--Salpeter equations beyond the ladder
approximation\cite{Bender:1996bb,Hellstern:1997nv,Bloch:1999vk} 
the absence of poles in the
quark propagator implies the absence of poles in the colour triplet $qq$
correlations as well. The latter correlations are the ones which we will call
diquarks in the following. It is interesting to note that diquarks are seen in
lattice calculations, see ref.\cite{Wetzorke:2000ez} and references therein.
One might even be puzzled by the fact that the corresponding spectral functions
for these confined objects are very similar to meson ones as can be seen in
fig.\ \ref{diq}.

\section{The relativistic Faddeev problem}

Equipped with models for the quark propagator we investigate the
relativistic three--quark problem. We neglect any three--particle irreducible
interaction graphs between the quarks, which defines the well--known Faddeev
problem.  For the two--quark correlations we use diquarks in the scalar and 
axialvector channel as discussed above, i.e., we use 
 a separable {\em ansatz} for the quark--quark
$t$--matrix of the form
\begin{equation}
t_{\alpha\gamma , \beta\delta}^{\hbox{\tiny sep}}(p,q,P) \, =
\chi_{\gamma\alpha}(p) \,D(P)\,\bar \chi_{\beta\delta}(q) \; +\;
\chi_{\gamma\alpha}^\mu(p) \,D^{\mu\nu}(P)
    \bar  \chi_{\beta\delta}^\nu (q)  \, .
\label{Gsep}
\end{equation}
Here, $P$ is the total momentum of the incoming and the outgoing quark-quark
pair, $p$ and $q$ are the relative momenta between the quarks.
$\chi_{\alpha\beta}(p)$ and $\chi_{\alpha\beta}^\mu(p)$ are
vertex functions of quarks with a scalar and an axialvector
diquark, respectively.
We parameterise the finite size of the diquark vertices by a dipole form.
We take the associated width parameter, that directly
influences the proton electric radius, to be of the order
300--400 MeV.
The inclusion of axialvector diquarks is the minimal
requirement to describe decuplet baryons and, see below, is
crucial for describing the nucleon electromagnetic form factors correctly.
The diquark propagators $D^{[\mu\nu]}$ are taken to
be free propagators of a spin--0 [1] particle multiplied by
the dressing functions defined in eqs.~(\ref{f1})~-~(\ref{f3}).

\begin{figure}
\centerline{\epsfig{file=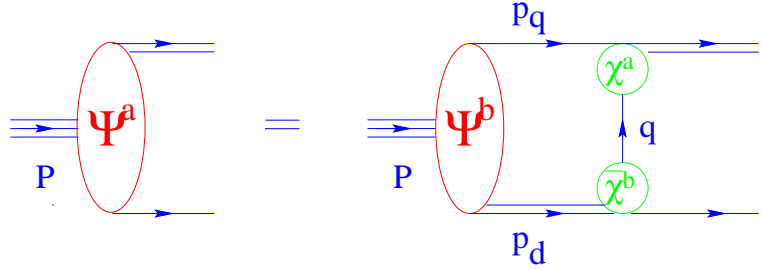, width=10cm}}
\caption{The Bethe--Salpeter equation for the spinorial baryon-quark-diquark 
wave function $\Psi$.}
\label{bse}
\end{figure}

Having imposed the separable {\em ansatz} (\ref{Gsep}) the Faddeev equations
reduce to a coupled system of Bethe--Salpeter equations describing baryons as
bound states of quarks and diquarks which interact by quark exchange. This
interaction is by virtue of the colour degree of freedom attractive. 
For the nucleon these equations read
\begin{eqnarray}
  \label{eq: BS_1}
  \int\frac{d^4p'}{(2\pi)^4} K(p,p',P)
  \pmatrix{\Psi^5 \cr \Psi^{\mu'}}(p',P) &=& 0 .
\end{eqnarray}
The interaction part of the kernel $K$,
\begin{eqnarray}
 K (p,p',P) &=& (2\pi)^4\: \delta(p-p')\: S^{-1}(p_q)
  \pmatrix{ D^{-1} & 0 \cr 0 & (D^{-1})^{\mu'\mu}} (p_d) + \nonumber \\
 & & \frac{1}{2}
  \pmatrix{\chi\: S^T(q)\:\bar\chi & -\sqrt{3} \chi^{\mu'}\: S^T(q)\:\bar\chi \cr
    -\sqrt{3}\chi\: S^T(q)\:\bar\chi^{\mu} &  -\chi^{\mu'}\: S^T(q)\:\bar\chi^\mu} \; ,
 \label{Gdef}
\end{eqnarray}
is given by the quark exchange, the least correlation required to restore
the Pauli principle for the three--quark state (for the definition of the 
involved momenta see Fig.~\ref{bse}). We have solved these equations
without further reduction and thus obtained fully Poincar{\'e}
covariant spinorial wave functions  $\Psi^{[\mu]}$.

\section{Electromagnetic form factors}

\begin{figure}[t]
\epsfxsize=0.8\linewidth
\centerline{\epsfbox{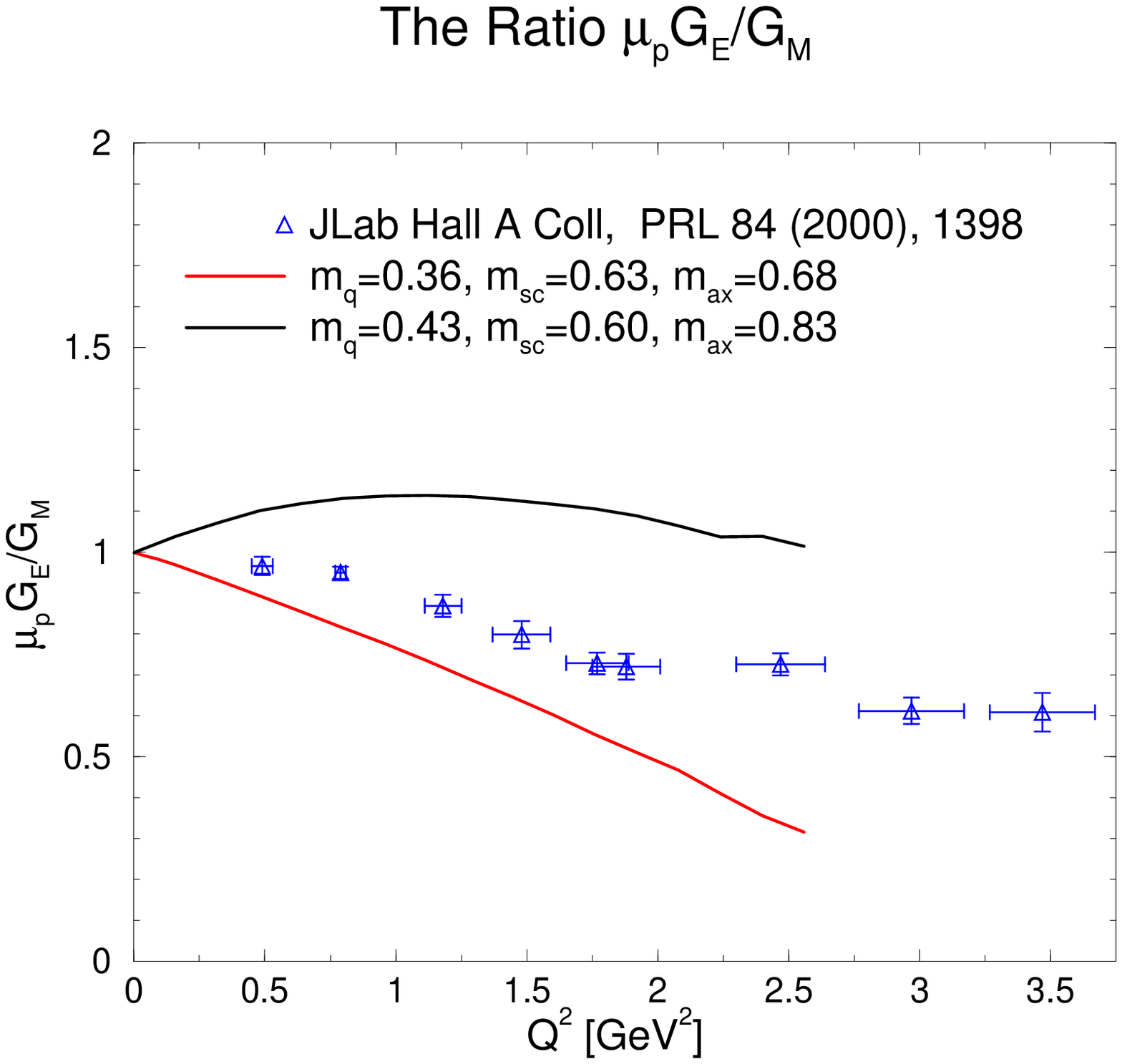} }
\caption{Ratio of electric to magnetic form factor of the proton as calculated
in our model with two parameter sets. $m_q$, $m_{sc}$ and $m_{ax}$ are the 
masses of quark, scalar and axialvector diquarks, respectively (in GeV).
The data are taken from ref.\protect\cite{Jones:2000rz}. 
\label{gem}}
\end{figure}

In a study employing free quark and diquark propagators\cite{Oettel:2000jj}
we calculated the nucleon electromagnetic form factors. Gauge invariance
and correct charge and norm were guaranteed by coupling the photon to
all possible places in the kernel of the Bethe--Salpeter equation
given in eq.~(\ref{Gdef}) \cite{Oettel:2000}. The results for the electric form factors
(up to momentum transfers of 2.5 GeV$^2$)
are in good agreement with the experimental data, nevertheless
it turned out  to be impossible to obtain a simultaneous correct description
of the nucleon magnetic moments and the mass of the $\Delta$ isobar.
Due to the free particle thresholds $m_q>411$ MeV had to be chosen
to obtain a bound $\Delta$ and these constituent quark masses
yielded proton magnetic moments $\mu_p \approx 1.9$. For lower
masses $m_q=360$ MeV we found $\mu_p \approx 2.5$, thus illustrating the
necessity to avoid the free--particle poles for the quarks.
We found that
20--25~\% axialvector correlations (measured by the ratio of the
norm contributions stemming from $\Psi$ and $\Psi^\mu$)
are needed to describe the ratio
of neutron to proton magnetic moment.
This conclusion is confirmed when looking at the ratio of electric to magnetic
form factor of the proton, see fig.\ \ref{gem}: The parameter set with 30~\%
axialvector correlations explains semi-quantitatively the data, the one with 
very small axialvector correlations not (for more details see 
ref.\cite{Oettel:2000jj}).

It is required by Ward identities to use dressed vertex functions when
employing the effective parameterisations of confinement (\ref{f1}) - 
(\ref{f3}), for details see ref.\cite{Ahlig:2000qu}. The resulting nucleon
magnetic moments are given in table \ref{tab1}: Their values have increased as
compared to ones calculated with free propagators, and they are as close to the
experimental values as can be reasonably expected. Please note that our model
does not contain effects like e.g.\ the pion cloud which certainly does give a
contribution to these quantities.  

\begin{table}[t]
\caption{Nucleon magnetic moments and e.m.\ radii with different confinement
parameterisations. The nucleon mass is used as input. The parameters are further
constrained by describing approximately the spectrum of octet hyperons 
(c.f.\ ref.\protect\cite{Oettel:1998bk}) as given here.
\label{tab1}}
\begin{center}
\footnotesize
\begin{tabular} {lrrrr} \hline \hline
 \\
   & Expon. & c.c.poles
   & Non-anal. & {Exp.} \\
\\ \hline \\
$\Lambda$ {\small [GeV]} & 1.13 & 1.13 & 1.12 & {1.12} \\
$\Sigma$ {\small [GeV]} & 1.30 & 1.22 & 1.21 & {1.19} \\
$\Xi$ {\small [GeV]} & 1.37 & 1.37 & 1.33 & {1.32} \\
\\ \hline \\
$\mu_p$  & 2.83 & 2.70 & 2.33 & {2.79} \\
$\mu_n$  & $-$2.37 & $-$2.08 & $-$1.82 & {$-$1.91} \\
$(r_{el})_p$ {\small [fm]} & 0.81 & 0.78 & 0.74 & {0.84} \\
$(r_{m})_p$ {\small [fm]} & 1.06 & 0.83 & 0.77 & {0.84} \\
\\ \hline \hline
\end{tabular}
\end{center}
\end{table}

We would like to mention that strong and weak form factors have been calculated
within this approach using free propagators\cite{Oettel:2000jj} and confined 
ones\cite{Ahlig:2000qu} (see also ref.\ \cite{Bloch:2000rm} for similar
calculations). Given all the calculations for static quantities and space-like
form factors one concludes that the use of confined propagators leads to
better results. However, they do not allow to distinguish between these
different parameterisations. Therefore we have to test the model propagators
at sufficiently large time--like momenta.

\section{Kaon photoproduction}

\begin{figure}[t]
\epsfxsize=0.8\linewidth
\centerline{\epsfbox{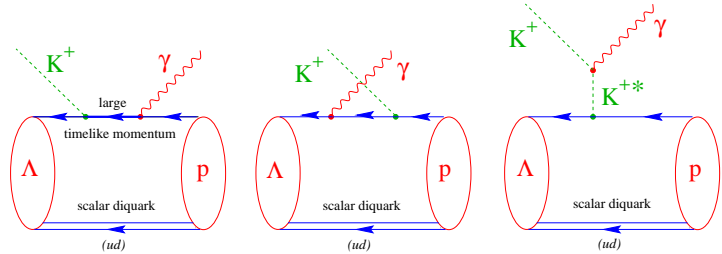} }
\caption{The three diagrams contributing mainly to kaon photoproduction in our
model. The right one tests the quark propagator at large time--like momenta.
The left one is a purely hadronic contribution. 
\label{pgkl}}
\end{figure}

Kaon photoproduction is a comparatively simple process in our model because
flavour algebra and parity dictates that in impulse approximation only diagrams
with scalar diquarks contribute, for a detailed discussions see
ref.\cite{Ahlig:2000qu}.  Comparison with data reveals that the exponential
propagator, see eq.\ (\ref{f2}), provides cross sections which are in
disagreement with data by orders of magnitude. Hereby the left-most diagram in
fig.~\ref{pgkl} overwhelms all other  contributions. For the
other two parametrisations the purely hadronic contribution, the  right-most 
diagram in fig.\ \ref{pgkl}, dominates. Given the theoretical problems with
non--analytic propagators\cite{Ahlig:2000qu} the one with complex--conjugated
poles, see eq.\ (\ref{f1}), provides therefore the best fit to experiment.

\section{Summary and Outlook}

We have described a model for baryons respecting full Poincar{\'e} invariance. 
This is possible by assuming separability in the
quark--quark $t$--matrix. This provides us with an effective definition for
extended diquarks. Baryons are then described by the solutions of a
Bethe--Salpeter equation. The binding mechanism is quark exchange: Due to colour
antisymmetry the Pauli principle leads to an attractive interaction.

We have modelled quark and diquark confinement via a parameterisation of
quark and diquark propagators. This has improved the results for static
quantities and space--like form factors of the nucleon. However, these
observables have not allowed to discriminate between different propagator types. 
This has been possible by
considering kaon photoproduction\cite{Ahlig:2000qu}: 
According to our analysis the propagators with 
complex conjugate poles are the best suited ones for future
investigations. Further insight into permissible dressing mechanisms will come 
from studies of nucleon structure functions which are under way\cite{Craig}.

\section*{Acknowledgements}
We thank Wally Melnitchouk, Andreas Schreiber, Peter Tandy, and Tony Thomas 
for organising this highly interesting workshop.

We are grateful to Steven Ahlig, Christian Fischer, Hugo Reinhardt, Craig
Roberts, Sebastian Schmidt,
Lorenz von Smekal and Herbert Weigel for their collaboration on related
issues in this model and helpful discussions. R.A.\ thanks Sebastian Schmidt
for support.

The work reported here has been supported by a Feodor--Lynen fellowship of
the Alexander-von-Humboldt foundation for M.O., by the 
DFG under contract nos.\ We 1254/4-2 and Schm 1342/3-1 and by COSY under
contract no.\ 4137660 .

\end{document}